\documentclass[conference]{IEEEtran}
\IEEEoverridecommandlockouts
\usepackage{cite}
\usepackage{url}
\usepackage{amsmath,amssymb,amsfonts}
\usepackage{algorithmic}
\usepackage{graphicx}
\usepackage{textcomp}
\usepackage{xcolor}
\usepackage{listings}
\usepackage[type={CC},modifier={by-nc-sa},version={4.0},]{doclicense} 

\makeatletter
\def\lst@makecaption{%
  \def\@captype{table}%
  \@makecaption
}
\makeatother
\def\BibTeX{{\rm B\kern-.05em{\sc i\kern-.025em b}\kern-.08em
    T\kern-.1667em\lower.7ex\hbox{E}\kern-.125emX}}

\begin{document}

\title{Towards a Hardware DSL Ecosystem: \\RubyRTL and Friends \thanks{\doclicenseText\doclicenseImage[imagewidth=6em]}}


\author{
    \IEEEauthorblockN{Jean-Christophe Le Lann}
    \IEEEauthorblockA{
        \textit{Labsticc UMR CNRS 6285} \\
        \textit{ENSTA Bretagne, France}\\
        \textit{jean-christophe.le\_lann@ensta-bretagne.fr}
    }
    \and
    \IEEEauthorblockN{ Hannah Badier}
    \IEEEauthorblockA{
        \textit{Labsticc UMR CNRS 6285} \\
        \textit{ENSTA Bretagne,France}\\
        \textit{hannah.badier@ensta-bretagne.fr}
    }
    \and
    \IEEEauthorblockN{Florent Kermarrec}
    \IEEEauthorblockA{
        \textit{Enjoy Digital} \\
        \textit{Landivisiau, France}\\
        \textit{florent@enjoy-digital.fr}
    }
}

\maketitle

\begin{abstract}
For several years, hardware design has been undergoing a surprising revival: fueled by open source initiatives, various tools and architectures have recently emerged. This resurgence also involves new hardware description languages. Inspired by the Migen Python community, we present RubyRTL, a novel internal domain-specific language for hardware design embedded in the Ruby language. Ruby -- which is best known in the field of web design -- has proven to be an excellent solution for the design of such DSLs, because of its meta-programming features. This paper presents the main aspects of RubyRTL, along with illustrating examples. We also propose a language-neutral interchange format, named Sexpir, that allows to seamlessly exchange RTL designs between Migen Python DSL and RubyRTL. This paves the way for interactions between various agile communities in the field of open source hardware design.
\end{abstract}


\section{Introduction}

For over four decades, EDA (Electronic Design Automation) has been relying on two main hardware description languages: VHDL and Verilog. These two languages were created in the 1980s, due to a clear need to formalize data exchanges between various companies involved in the exponential growth of micro-electronics \cite{asv_tides_2003}. However, their use was in the end not exclusive to the documentation of engineering activities: they are nowadays also used for simulation --at various level of abstraction-- and for the actual synthesis on silicon. More precisely, these two HDLs are mainly used for RTL design entry, down to logic synthesis. Despite very high expectations, intense research and tremendous results, high-level synthesis (HLS) -- which consists in generating a RTL design from a sequential algorithmic description -- has not replaced pure RTL design activities. Indeed, many components of complex system-on-chips (SoCs) are not particularly amenable to algorithmic descriptions, and require instead a careful handcrafted elaboration at the register-transfer level. This ongoing reliance on RTL HDLs instead of HLS could be considered disappointing. Instead, we argue here that this opens up unprecedented ways of thinking and designing hardware.

The revival of RTL design methods is led prominently by the use of classic mainstream languages. They differ greatly from the previous VHDL and Verilog approaches, whose use was often limited to relatively few specialists. By using Python, Haskell, C++ or Scala, RTL design is now within the reach of a wide range of engineers. The integration of RTL concepts in these classic languages --either through libraries, design patterns or even full hardware-oriented compilers-- has given rise to the notion of internal RTL DSLs (domain-specific languages). We believe that this trend is not a passing trend: relying on classic languages has already proven to be a stimulating way to bring new ideas to the forefront, but more importantly an effective method to design real systems.

In this paper, we propose a new RTL DSL based on Ruby, a object-oriented language (not to be mistaken for the prototype language Ruby, used in formal verification \cite{jones1990circuit}). Ruby was born in Japan in the mid 1990's. Its creator --Yukihiro Matsumoto-- was strongly influenced by the elegance of Smalltalk and Lisp, and still insists on the need to bring "joy and fun" to programmers. The Ruby language has become known worldwide thanks to the Ruby-on-Rails framework, which uses meta-programming intensively. The framework is often presented as a set of DSLs dedicated to the web. Indeed, Ruby is particularly well suited for meta-programming activities: while often considered exotic in other programming languages, meta-programming and introspection are encouraged in Ruby. For instance, beginners in Ruby resort to meta-programming to create instance variable accessors : getters and setters methods are created dynamically. The call for this creation (attr\_accessor) may be mistaken for a Ruby keyword, but is rather a class-level method call. Such illusions are mainstream in Ruby. This probably influenced Dave Thomas to state: "programming is the creation of your own DSL". Our work on RubyRTL leverages these capabilities of Ruby for hardware design.

The rest of this paper is structured as follows: next section presents some background material and related work. Section III presents RubyRTL DSL itself: both its syntax and internal design are exposed, illustrated via simple code examples. Section IV goes a step further by providing a mean to exchange IPs between RubyRTL and Python Migen DSL, back and forth. Section V concludes this work-in-progress. All the artefacts used in this paper are available on Github and Rubygems websites as RubyRTL and Sexpir under MIT permissive licence.

\section{Related work}
The idea of resorting to various DSLs for FPGA design is now a reality: in \cite{Kapre_fpl16}, the authors proposed a survey of this trend. Their paper lists several DSLs dedicated to various ESL (electronic system level) and EDA activities. Our work focuses on RTL design only. MyHDL \cite{DBLP:conf/fpga/JaicS15}, but even more Migen Python DSL \cite{migen}, are the main source of inspiration for RubyRTL. Migen approach enforces the synchronous design paradigm, by a straightforward description of combinatorial and sequential assignments. This allows to get rid of most problematic event-driven aspects of VHDL and Verilog, which can lead to synthesis-simulation mismatches. Such a DSL is sometimes called "hardware construction language". Migen has proven to be a solid and effective tool for FPGA design. For instance, LiteX IP library is entirely described in Migen \cite{kermarrec2019litex}. LiteX has also proven effective in that it allows to describe full SoC platforms: LiteX supports various externally described softcore CPUs, including LM32, OpenRISC, OpenPower Microwatt and several RISC-V cores like picorv32 and VexRiscv. Python is also used for some other experimental hardware DSLs \cite{clow_fpl17}, standing as inputs of {\it hardware generators}, even for analog purposes \cite{nikolic_esscir_18}.

However, the RTL DSL trend is not limited to Python: first, we need to mention SystemC, which was among the firsts to adopt such an approach (embedded in C++), while also offering industrial-level support. SystemC however replicates the event-driven paradigm in terms of simulation model. More recently, Scala and Haskell \cite{DBLP:conf/codes/ZhaiTLKE15} have also drawn increasing attention. The Scala-based Chisel DSL \cite{chisel_12} has contributed significantly to the joint emergence of RISC-V community.  It is closely associated to Firrtl intermediate representation \cite{Izraelevitz_iccad_17}. SpinalHDL, its direct competitor, demonstrates an even faster progress and adoption \cite{spinalHDL}. The Haskell approaches take advantage of the functional nature of the language. Functional languages map well to hardware: immutable data structures and the absence of side-effects make formal reasoning about programs easier. Even more importantly, functional languages are inherently concurrent and race-free. In any case, the goals are twofold: increasing the level of abstraction and offering the comfort of a classic, well-known language as DSL host.

\section{RubyRTL}

RubyRTL \cite{ruby_rtl}  aims at offering a straightforward way to describe hardware at the RTL level. In this section, we first present the design of the DSL itself. Then, we provide illustration of the DSL syntax through some basic examples.

\subsection{DSL internals}
We leverage the exceptional capability of Ruby in terms of embedding of internal DSLs. It needs to be noted however that, similarly to Python Migen, the real efficiency of RubyRTL comes from the interleaving of classic Ruby code and its embedded DSL:  parameterization and generic instanciations are notably handled by the host language easier than in classic HDLs.  As far as the DSL is concerned, we repeatedly resort to Ruby syntactical capacity to give the user the illusion of adding new keywords to the language. This illusion is essentially based on three syntactical tricks:
\begin{itemize}
\item Ruby method calls do not require parentheses. For instance, in RubyDSL \textbf{input :a} is a method call that reflects the idea of having a dedicated keyword \texttt{input}, {\it followed by} a name \texttt{a} (\textbf{:a} is a Ruby Symbol which can be seen as an immutable String).
\item Use of hash element association symbol '\texttt{=>}' to suggest the idea of a mapping. For instance, when we need an explicit type name, we can write: \\
\texttt{input :a => :complex}.
Here, we have a method call (input) with a Hash pair as argument.
\item Extensive use of Ruby Blocks (known as {\it closures} in other languages), delimited by braces '\texttt{\{...\}}' and passed as method arguments to mimic the nesting of code. For instance, to provide the illusion of having an "If", "Then", "Else" statement, we have written three methods named \texttt{If, Then and Else} (capitals are compulsory in order to avoid collision with classic Ruby statements). The If method is given a DSL expression as first argument, while the second, representing the body of the If, is a Ruby block. etc.
\end{itemize}
During the interpretation of Ruby, the DSL objects build their own abstract syntax tree (AST) nodes internally. An object named Compiler is responsible for the elaboration of the complete AST.

\subsection{DSL Syntax}
The new RTL-oriented keywords embedded in RubyRTL are directly inspired by designers best practices. For instance, while neither VHDL nor Verilog propose dedicated keywords for finite-state machine concepts (fsm, state, transition), we have created these notions directly in the DSL (as a remainder, in VHDL and Verilog, designers use a coding template made of assignments and switch-case mechanism to model FSMs).

\begin{figure}[htbp]
\centerline{\includegraphics[scale=0.28]{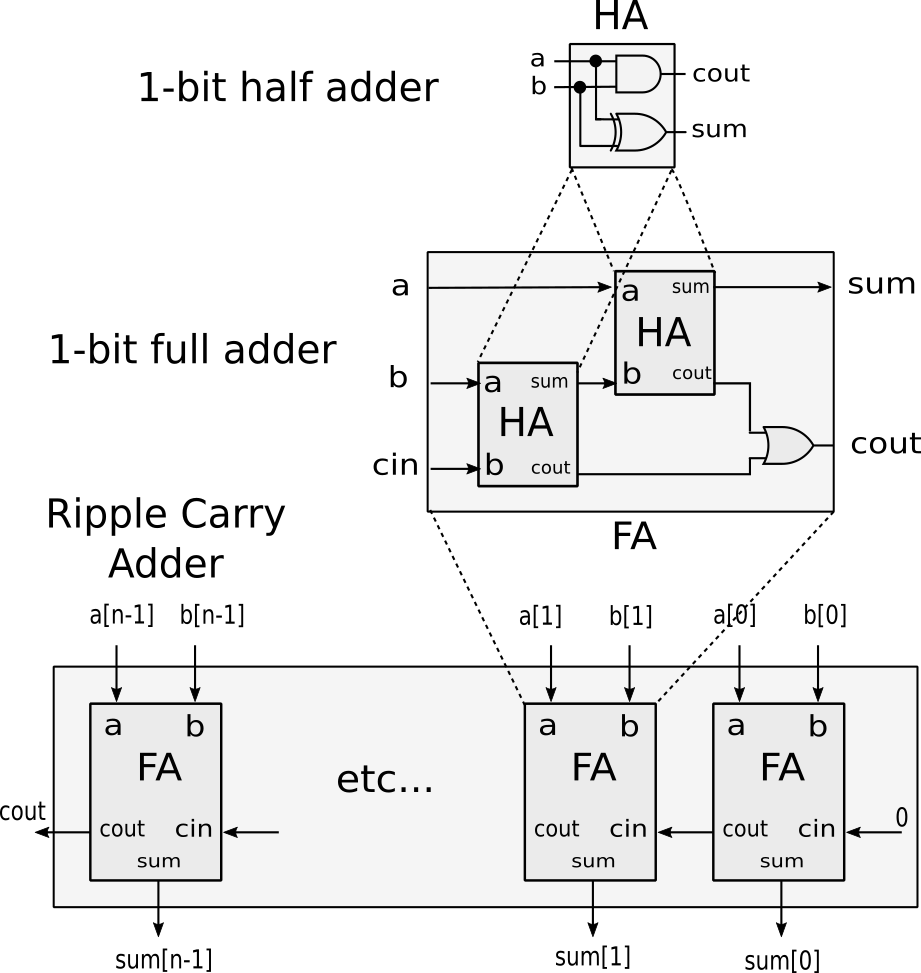}}
\caption{Hierarchical design of a ripple-carry adder (recall)}
\label{fig_adder}
\end{figure}

Below, we review some of RubyRTL's main syntactical features. Please note that for sake of brevity, the call to the Ruby library (gem) and the inclusion of the RubyRTL module, which always account for 2 additional lines of code at the beginning of the file, are not included in the following descriptions. We first start by the "Hello world" of logic design: the hierarchical design of a generic ripple-carry adder. The architecture of this circuit is shown on fig \ref{fig_adder}.

\subsubsection{Inputs, Outputs and Assignments}
The following example describes an elementary half-adder (listing \ref{lst1:mxm}): it takes two bits as inputs and sums them, possibly requiring a carry signal. Two gates (and, xor) are necessary. The syntax of inputs and outputs was explained in the previous section. By default, when no type is associated with the port name, a type Bit is assumed. The syntax of assignments is remarkably clear: it relies on a method call, named 'assign', that receives a Ruby less-or-equal comparison as argument. RubyRTL mechanisms then transform the whole as an assignment AST node. The magic of elaborating an AST using an internal DSL also resides in the fact that all references are resolved easily, without the hassle of maintaining a symbol table. For instance, in the half-adder, the apparent reference to input 'a' in the first and second assignment relates to a getter method named 'a' that has been generated by meta-programming during the input method call.

\begin{lstlisting}[
  caption=Half Adder RubyRTL code,
  label=lst1:mxm,
]
class HalfAdder < Circuit
  def initialize
    input :a,:b
    output :sum,:cout

    assign(sum   <= a ^ b)
    assign(cout  <= a & b)
  end
end
\end{lstlisting}

\subsubsection{Structural descriptions}
The second example (listing \ref{lst2:mxm}) aims at describing a classic 1-bit full-adder, which adds two bits and an input carry and returns the sum and an input carry. The circuit is designed with reuse in mind: the full-adder can be described using the preceding half-adder. We first \texttt{require} the preceding Ruby code. In the initialize method (sort of constructor), we mimic the instanciation of two half adders using a DSL keyword 'component' (either via true instanciation of a HalfAdder object or the class itself, as commented in the code). The assignments now allow to access the instances ports via a pointed notation.

\begin{lstlisting}[
  caption=Full Adder RubyRTL code,
  label=lst2:mxm,
]
require_relative 'half_adder'

class FullAdder < Circuit
  def initialize
    input  :a,:b,:cin
    output :sum,:cout

    component :ha1 => HalfAdder     # class...
    component :ha2 => HalfAdder.new # or ...obj

    assign(ha1.a <= a )
    assign(ha1.b <= b )
    assign(ha2.a <= cin)
    assign(ha2.b <= ha1.sum)
    assign(sum   <= ha2.sum)
    assign(cout  <= ha1.cout | ha2.cout)
  end
end
\end{lstlisting}

Now equipped with a one bit full-adder, we can describe a generic arithmetic adder operating on two n-bits operands (listing \ref{lst3:mxm}). This introductory-level exercise about genericity in Verilog or VHDL is even more easily solved using RubyRTL. We interleave classic Ruby statements for loops and DSL assignments to assemble the n full-adder components. They are first pushed into a Ruby array (adders) and then referenced by their index.

\begin{lstlisting}[
caption=generic n-bits ripple-carry adder RubyRTL code,
  label=lst3:mxm
]
class Adder < Circuit

  def initialize nbits
    input  :a    => nbits
    input  :b    => nbits
    output :sum  => nbits
    output :cout

    # create  components
    adders=[]
    for i in 0..nbits-1
      adders << component("fa_#{i}"=>FullAdder)
    end

    # connect everything
    for i in 0..nbits-1
      assign(adders[i].a <= a[i])
      assign(adders[i].b <= b[i])
      if i==0
        assign(adders[0].cin <= 0)
      else
        assign(adders[i].cin<=adders[i-1].cout)
      end
      # final sum
      assign(sum[i]        <= adders[i].sum)
    end
  end
end
\end{lstlisting}

\subsubsection{Behavioral descriptions}

RubyRTL also offers means to describe so-called "behavioral" RTL statements. Listing \ref{lst4:mxm}) illustrates the DSL syntax of a counter that counts from 0 to 1 at each tick. This sequential behavior is made explicit by using the \texttt{sequential} DSL keyword. Same as Migen, an implicit clock is assumed. The current version of RubyRTL also assumes an asynchronous and synchronous reset for every D flip-flop. This example makes use of an \texttt{If...Else} statement: again we can recall that we have instead two method calls (If and Else). During elaboration, the two resulting separated sub-nodes, representing the two branches, are merged appropriately. This was the Migen designers choice, who instead relied on pointed notation for the Else part. We believe that RubyRTL syntax is here more elegant and natural. Our Github repository gives some other examples: \texttt{Case..when..default} statement is also available.

\begin{lstlisting}[
caption=Counter in RubyRTL code illustrating If/else syntax and sequential process,
  label=lst4:mxm
]
class Counter < Circuit
  def initialize
    input  :tick
    output :count => :byte

    sequential(:counting){
      If(tick==1){
        If(count==255){
          assign(count <= 0)
        }
        Else{
          assign(count <= count + 1)
        }
      }
    }
  end
end
\end{lstlisting}

\subsubsection{Finite state machines}
Finally, FSMs are also easily described in RubyRTL. The same mechanism of method call syntax with block closures as argument is used, with a deeper nesting. Notice that by default, the first state is the reset state. By default as well, all assignments in the state machines are considered synchronous. Another keyword \texttt{comb\_assign} (not used here) can be used to describe a combinatorial assignment within a state, if needed.

\begin{lstlisting}[
caption=generic n-bits ripple-carry adder RubyRTL code,
  label=lst5:mxm
]
class FSM1 < Circuit
  def initialize
    input :go
    output :f => :bv2
    fsm(:simple){
      assign(f <= 0)
      state(:s0){
        assign(f <= 1)
        If(go==1){
          next_state :s1
        }
      }
      state(:s1){
        assign(f <= 2)
        next_state :s2
      }
      state(:s2){
        assign(f <= 3)
        next_state :s0
      }
    }
  end
end
\end{lstlisting}

\subsection{Declaring complex data types}
RTL design in 2020 is highly structured, and makes use of complex data types. RubyRTL aims at providing similar capabilities to VHDL in terms of data types creation. We have a \texttt{typedef} DSL keyword, as well as Record and Array keywords. The example shows the creation of two new types and their use for the description of an array of 256 complex numbers, each coded as int6 integers. If declared before any inputs or outputs, the declared types are generated in VHDL in a dedicated package that allows to refer to these types within the VHDL entity declaration.

\begin{lstlisting}[
caption=Arrays and Record type definition in RubyRTL code,
  label=lst6:mxm
]
typedef :cplx => Record(:re=>:int6,:im=>:int6)
typedef :cplx_ary => Array(256,:cplx)
wire :mem => :cplx_ary
for i in 0..255
   assign(mem <= {re: i, im: i*2})
end
puts mem[13][:im] # prints 26
\end{lstlisting}

\subsubsection{Automatic cast and type conversions}
We designed RubyRTL with VHDL in mind. We wanted to simplify the burdens of type castings and conversions, which are particularly verbose in VHDL and can be surprising for novice designers. We experiment with various automatic type conversions that will make sense to most designers: we followed the principle of least surprise (also common in the Ruby community) during the expression of RubyDSL circuits. The RubyRTL contextual analyzer automatically applied such transformations. Some examples are given in the next listing. One of the most impacting choices is the conversion of single bit signal (eg. a start or stop signal) to an unsigned integer of size 1, when such signal is compared to 1 (or 0): in that case, we assume the designer prefers writing such 1 (or 0) as a plain Ruby Integer literal 1 or 0, instead of a more complex expression indicating the single bit nature of the signal. This type of conversion can be examined in the VHDL code provided in the next subsection: the 'go' signal of the FSM, when compared effectively to a Ruby integer 1, is first translated to such an unsigned integer.

\begin{lstlisting}[
caption=Experimental automatic type conversion,
  label=lst7:mxm
]
wire  :a,:f1,:f2
wire  :b => :bv8
wire :w1 => :bv8
wire :w2 => :bit
wire :w3 => :int8
assign(f2 <= 1)
# bit <= ruint1   ===> bit <= bit
assign(f1 <= 42)
# bit <= ruint6   ===> ERROR
assign(w1 <= a + 1)
# bv8  <= bit + ruint1 ==> bv8 <= resize(bit,8)
assign(w2 <= a + 1)
# bit  <= bit + ruint1   ==> ERROR
assign(w2 <= 1 + 1)
# bit  <= ruint2  !!!    ==> ERROR
assign(w3 <= a + 5)
# int8 <= bit + ruint3   ==> int8 <= signed(resize(ruint3,8))
\end{lstlisting}

\subsection{Code generation}
RubyRTL is designed such that every construct of the DSL is synthesizable. The object DSL compiler provided in RubyRTL tries to generate VHDL code by default. It also offers additional services, such as AST viewing (Graphviz dot file), pretty printing and finally VHDL code generation. All theses compiler passes, designed using a Visitor pattern, operate on the AST of the circuit object. The VHDL code of the FSM described earlier is given as generated code example here. It should be noted that we rely on a single process state machine template, which can easily be modified, if needed.

\lstset{
  frame=single,
  language=VHDL,
  basicstyle=\small,
}
\begin{lstlisting}[
caption=FSM VHDL code generated,
  label=lst8:mxm
]
// skipped headers and entity...
architecture rtl of fsm1_c is
  type simple_state_t is (s0,s1,s2);
  signal simple_state : simple_state_t;
begin
  simple_update : process(reset_n,clk)
  begin
    if reset_n='0' then
      simple_state <= s0;
      f <= (others=>'0');
    elsif rising_edge(clk) then
      if sreset='1' then
        simple_state <= s0;
        f <= (others=>'0');
      else
        case simple_state is
          when s0 =>
            f <= to_bv(1,2);
            if (to_uint(go,1) = 1) then
              simple_state <= s1;
            end if;
          when s1 =>
            f <= to_bv(2,2);
            simple_state <= s2;
          -- skipped for paper brevity
          when others =>
            null;
        end case;
      end if;
    end if;
  end process;

\end{lstlisting}

\section{Towards IP interchange between Migen and RubyRTL: early experiment}

\begin{figure*}[tbp]
    \centerline{\includegraphics[scale=0.08]{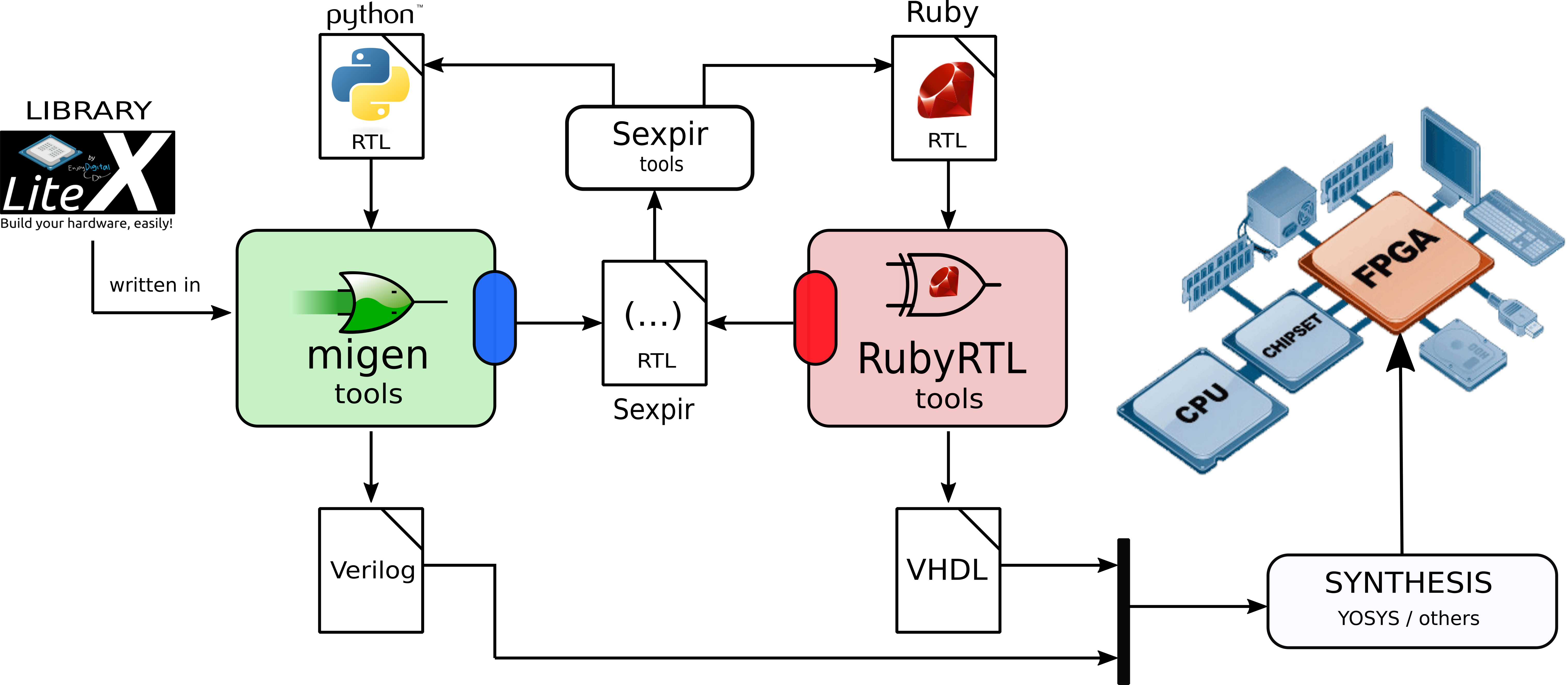}}
    \caption{Experimental toolchain for Migen-RubyRTL IP interchange}
    \label{toolchain}
\end{figure*}
In this section, we automatically translate an IP described in Migen to RubyRTL, and then generate VHDL code. This is especially useful since Migen only provides Verilog generation. Using a symmetrical experiment, we push the idea of a full IP interchange between the two DSLs. To proceed step-wise, we have designed an intermediate representation, named Sexpir. The association of these three tools (Migen, Sexpir and RubyRTL) results in an interesting toolchain (fig \ref{toolchain}), capable of handling different uses cases, discussed in the sequel. The following section presents Sexpir, use cases, and an early experiment with this toolchain.

\subsection{Sexpir intermediate representation}
 Sexpir \cite{sexpir} textual representation and tooling was prototyped to ease the interchange between Migen and RubyRTL. As its name suggests, Sexpir is based on s-expressions, which are language neutral, friendly and easily parsed in several programming languages.  Sexpir is voluntarily much more modest than other intermediate representations such as FIRRTL \cite{Izraelevitz_iccad_17}. We purposely restrict Sexpir to a strict minimum so that it can represent the same hardware concepts as manipulated in Migen and RubyRTL. Sexpir does not handle complex transformations, like intended by Firrtl tooling. In particular, behavioral constructs are preserved: RTL netlist inference is deferred to either Migen, RubyRTL, or any new framework that is able to load Sexpir files.
\lstset{
  frame=single,
  language=lisp,
  morekeywords={circuit,input,name,type,signal,assign,bits_sign,combinatorial,case,when,if,then},
  basicstyle=\small,
}
\begin{lstlisting}[
caption=Sexpir code (excerpt) generated by an experimental Migen code generation pass,
  label=lst9:mxm
]
(circuit uart
  (input (name sys_rst) (type bv1))
  (input (name sys_clk) (type bv1))
  (signal (name rx_data]) (bits_sign 8))
  ;; ...skipped
  (assign rx_strobe (== rx_counter 0))

  (combinatorial nil
    (assign rx_error 0)
    (assign fsm0_next_state 0)
    ;; ...skipped
    (case fsm0_state
        ;; ...skipped
      (when 2
        (if (== rx_strobe 1)
          (then
            ;; ...skipped
          )
        )
      )
      ;; skipped
    )
    ;; ...skipped
  )
)
\end{lstlisting}

\subsection{Use cases}
Various use cases for RubyRTL are envisioned here. As stated in this paper, the existence of RubyRTL is first and foremost beneficial for the Ruby community itself. However, with VHDL as a first code generation target, RubyRTL can be envisioned as a way to prototyping VHDL-oriented designs in a concise and natural way. In this regard, our future work will consist in providing various RubyRTL facilities for such users: high-speed cycle-based simulation, RTL basic synthesis and RTL netlist visualisation and animation. Another use case relates to Sexpir IR, which allows to extend the ecosystem with other DSLs. Finally, the whole toolchain should  allow to translate complex libraries like LiteX into VHDL.

 \subsection{Experiment}
To test the complete toolchain, we took an UART IP \cite{uart} designed by Whitequark in both Migen and handwritten Verilog. We modified Migen to add a Sexpir generation step from Migen post-elaboration representation: it intervenes just before Verilog code generation. This choice may be discussed and revised: some syntax constructs (like FSM) seem lost during this Migen process, but the whole RTL code is preserved. The Sexpir compiler then generates RubyRTL code, which in turn generates VHDL that can be simulated using for example GHDL \cite{ghdl}. Prior to code generation however, Sexpir compiler needs to elaborate a data dependency graph to recover inputs and outputs, initially marked as plain signals in both Migen and Sexpir files.
 This experiment makes us very confident for more extensive automatic IP interchange between the two DSLs : we plan to automate the translation of the entire LiteX library using the same procedure. Future work will also consist of analyzing the impact of new versions of Migen (nmigen) on the toolchain, and adapting it accordingly.

\section{Conclusion and future work}
In this paper, RubyRTL domain-specific language was proposed. RubyRTL contributes to the expansion of the hardware design community, by allowing Ruby programmers to tackle FPGA and ASIC design. We hope that its simplicity and direct exposure of RTL concepts may also invite newcomers to consider RubyRTL as a premier language for open source hardware prototyping. Our paper also proposed a second tool, named sexpir, which acts as a gateway for language-neutral IP interchange.


\bibliographystyle{IEEEtran}
\bibliography{biblio}

\end{document}